\documentclass[aps,pra,twocolumn,showpacs,amsmath,amsmath,amssymb,amsfonts]{revtex4-1}
\usepackage{graphicx}
\usepackage{dcolumn}
\usepackage{bm}
\usepackage{color}
\usepackage{multirow}
\usepackage{bbold}

\begin{document}
\author{Cenap Ates}
\author{Beatriz Olmos}
\author{Juan P. Garrahan}
\author{Igor Lesanovsky}
\affiliation{School
of Physics and Astronomy, The University of Nottingham, Nottingham,
NG7 2RD, United Kingdom}
\title{Dynamical phases and intermittency of the dissipative quantum Ising model}
\date{\today}
\begin{abstract}
We employ the concept of a dynamical, activity order parameter to study the Ising model in a transverse magnetic field coupled to a Markovian bath. For a certain range of values of the spin-spin coupling, magnetic field and dissipation rate, we identify a first order dynamical phase transition between active and inactive {\em dynamical phases}. We demonstrate that dynamical phase-coexistence becomes manifest in an intermittent behavior of the bath quanta emission. Moreover, we establish the connection between the dynamical order parameter that quantifies the activity, and the longitudinal magnetization that serves as static order parameter. The system that we consider can be implemented in current experiments with Rydberg atoms and trapped ions.
\end{abstract}


\maketitle

\section{Introduction}
The remarkable progress in the control of ultra cold atomic gases and trapped ions has opened a new door for studying  dissipative many-body quantum systems. Very recently it was shown that a carefully designed dissipative dynamics arising from an engineered heat bath can lead to the formation of pure and coherent many-body quantum states \cite{dimi+:08,vewo+:09,diyi+:10,wemu+:10,bamu+:11}. Moreover, it was demonstrated that open many-body systems with competing dissipative and coherent interactions possess a rich phase structure \cite{wevo+:05,syba+:08,dito+:10,spst+:10,todi+:11,leha+:11,leha+:12}. However, despite their intrinsic dynamical nature, these phases are often classified by means of equilibrium order parameters, such as particle densities or spatial correlation functions.

In this work we pursue a complementary route which aims at describing dynamical phases in terms of strictly dynamical order parameters. The approach that we will use is called \emph{thermodynamics of trajectories}. It has already proven to be useful for the study of classical many-body systems displaying complex cooperative dynamics such as glasses \cite{gale+:07,heja+:09}. Recent theoretical work \cite{gale:10} has adapted this approach for simple open quantum systems and has shown that dynamical phase behavior can be uncovered by means of an \emph{activity} order parameter \cite{leap+:07,gale+:07,mane:08}, i.e. an observable that counts the emission of quanta from an open quantum system into its environment (events often referred to as quantum jumps). 

\begin{figure}
\centering
\includegraphics[width=\columnwidth]{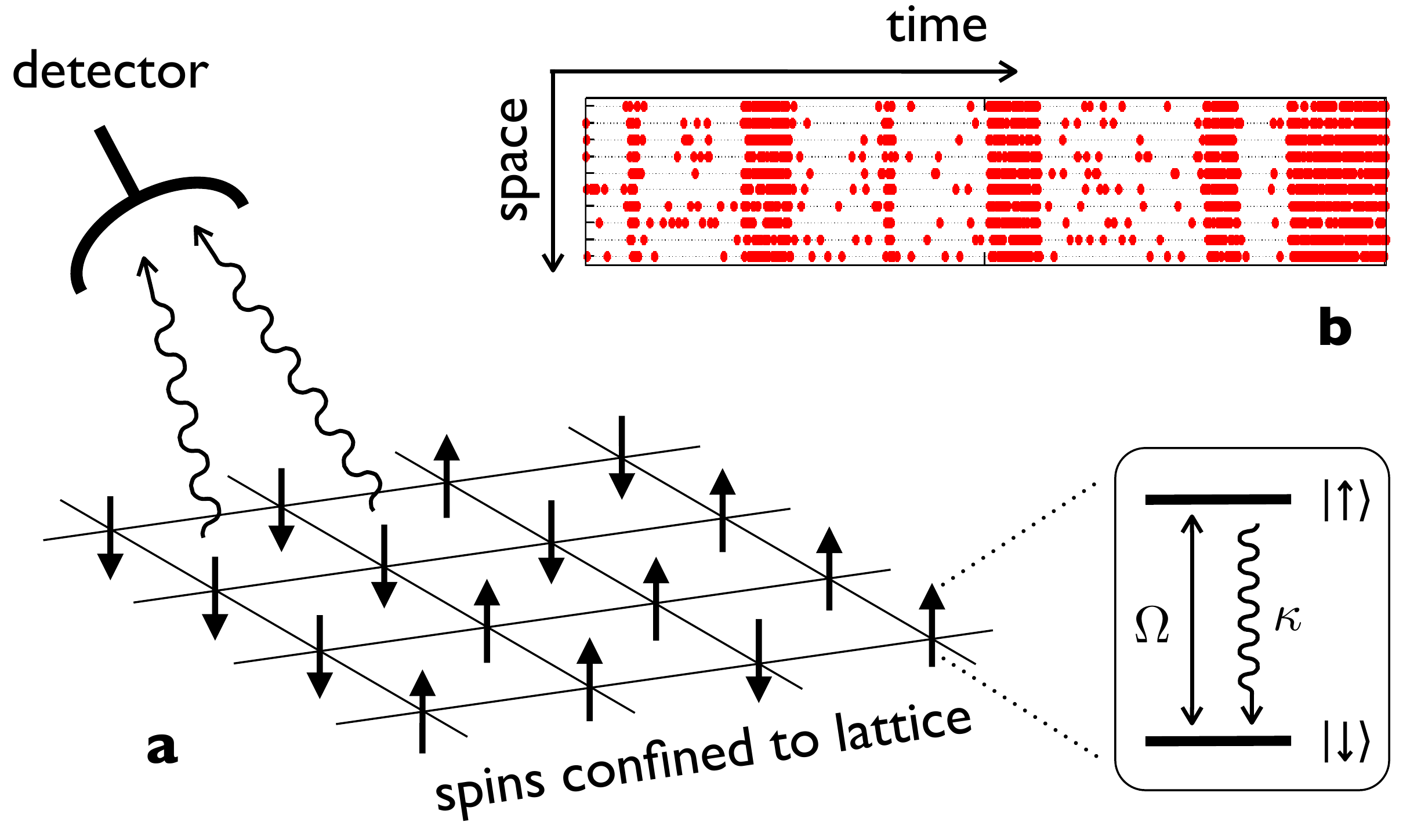}
\caption{(Color online) (a) Schematic of the open Ising model in a transverse field. Internal states of trapped atoms or ions are described by spin-$1/2$ degrees of freedom on a lattice. The interaction between nearest neighbors has strength $V$, and there is an applied transverse field $\Omega$ (e.g. the Rabi frequency of a laser). Interaction with the radiation field leads to the incoherent emission of bath quanta (e.g. photons) with rate $\kappa$, which are detected and counted. (b) Quantum jump trajectory in a one-dimensional version of (a). Emitted photons are temporally (and spatially) resolved, so each point indicates where and when a quantum jump event took place. This particular trajectory shows intermittency, which manifests the coexistence of an active and an inactive dynamical phase (see text).}
\label{fig:system}
\end{figure}

Here, we show that the thermodynamics of trajectories approach can also be used to gain insights into the dynamical behavior of interacting many-body quantum systems. We illustrate this by studying a quantum Ising model in a transverse magnetic field subject to Markovian dissipation that couples to individual spins (cf.\ Fig.\ \ref{fig:system}). Beyond the fact that Ising models like the one we study here serve as paradigmatic examples of many-body systems, the system at hand can be implemented with recently developed techniques in experiments with trapped ions \cite{bamu+:11,scpo+:11} or Rydberg atoms \cite{sawa+:10,viba+:11,leha+:12}.

We show that depending on the experimental parameters the system is either found in a specific dynamical phase or at coexistence conditions of two such phases. We demonstrate that this phase coexistence is accompanied by strong fluctuations in the activity that become manifest in pronounced intermittency in the emission of the bath quanta. The thermodynamics of trajectories approach also sheds light on the intermittency that has recently been theoretically found in an open fully connected spin model \cite{leha+:12}.

The paper is organized as follows. In Section \ref{sec:method} we outline the thermodynamics of trajectories approach for quantum systems. In particular, we establish the relation between the dynamical phases, as classified by their activity, and equilibrium phases characterized by static observables (Sec.\ \ref{sec:connection}). In addition, we show that such a direct connection does, in general, not hold between dynamical and static fluctuations. Following these general considerations, we study the dynamical behavior of the dissipative quantum Ising model in a transverse field using the thermodynamics of trajectories approach in Section \ref{sec:oising}. After briefly discussing a possible realization of this model with highly excited atoms (Sec.\ \ref{sec:experiment}) we investigate the dynamics of the system on the level of mean-field theory (Sec.\ \ref{sec:meanfield}) as well as by using the numerical techniques of exact diagonalization (Sec.\ \ref{sec:diagonalization}) and Quantum Jump Monte-Carlo simulations (Sec.\ \ref{sec:montecarlo}). We focus, in particular, on the regime in which the model displays strongly intermittent behavior in the emission of bath quanta. We show that this is a consequence of the system being at a \emph{first order coexistence point between two dynamical phases}.

\section{Thermodynamics of trajectories}\label{sec:thermo_of_trajec}
\subsection{Dynamical order parameter and ensembles of quantum trajectories}\label{sec:method}
Let us consider a quantum many-body system composed of $N$ particles coupled to a Markovian bath. The state of the system is described by the density matrix $\rho$ and its time-evolution governed by the Master equation $\partial_t \rho = \mathcal{W} (\rho)$ with super-operator,
\begin{align}
 \mathcal{W}(\bullet)=-i\left[H,\bullet\right]+\sum_{\nu=1}^N J_{\nu} \bullet J_{\nu}^\dagger-\frac{1}{2}\sum_{\nu=1}^N\left\{J_{\nu}^\dagger J_{\nu},\bullet\right\},
 \label{eq:ME}
\end{align}
and system Hamiltonian $H$  \cite{brpe:02}. The coupling to the thermal bath is assumed to take place at the single particle level, and is mediated by the Lindblad operators $J_{\nu}$ \cite{brpe:02}. These operators induce sudden changes of the state of the system and, therefore, describe spontaneous emission events of quanta into the bath.

We are interested in the statistics of these emission events and their connection to the quantum dynamics of the system.
To illuminate this we employ the thermodynamics of trajectories approach, which will allow us to gain immediate insights into the emission characteristics of bath quanta. The idea, which for quantum systems is outlined in Ref.\ \cite{gale:10}, is to consider time records of emission events ---so-called quantum jump trajectories--- and to study ensembles of trajectories in a manner analogous to how one studies ensembles of microstates (or configurations) in equilibrium statistical mechanics. A trajectory of length (i.e. observation time) $t$ contains $K$ emission events; see e.g.\ Fig.\ \ref{fig:system}(b). The probability for such a trajectory to occur is given by $p_K(t)=\mathrm{Tr}\, \rho^{(K)}(t)$ where $\rho^{(K)}(t)$ is the projection of the density matrix $\rho$ onto the subspace in which $K$ emission events have taken place \cite{zoma+:87}. The generating function of the probability distribution $p_K(t)$ is defined as
\begin{eqnarray}
  Z&=&\sum_{K=0}^\infty e^{-sK}p_K(t)=\sum_{K=0}^\infty e^{-sK}\mathrm{Tr}\,\rho^{(K)}(t) \nonumber \\
  &=&\mathrm{Tr}\sum_{K=0}^\infty e^{-sK}\rho^{(K)}(t) \equiv \mathrm{Tr}\, \rho_s(t).
\end{eqnarray}
Here $\rho_s(t)$ is the Laplace transformed density matrix which evolves according to the generalized Master equation $\partial_t\rho_s =\mathcal{W}_s(\rho_s)=\mathcal{W}(\rho_s)+\mathcal{V}_s(\rho_s)$ \cite{zhbr+:03,esha+:09} where
\begin{eqnarray*}
  \mathcal{V}_s(\bullet)=(e^{-s}-1)\sum_{\nu} J_{\nu} \bullet J_{\nu}^\dagger.
\end{eqnarray*}
This generalized Master equation is not probability conserving, and corresponds to non-physical dynamics where the probability of quantum jump trajectories with $K$ events are biased by a factor $e^{-sK}$. This means that for negative (positive) $s$ trajectories with more (less) emission events than the average are more likely to occur. The physical (probability conserving) dynamics takes place at $s=0$, where the generalized Master operator coincides with (\ref{eq:ME}). However, as we show below, information about the behavior of the system in the vicinity of this physical point (i.e. at $s \neq 0$) can be crucial for the understanding of its emission dynamics \cite{gale:10,bu:10}.

We are interested in dynamical properties in the stationary regime. Therefore, we consider observation times long enough for all initial transient effects to have become negligible. For these long times the generating function $Z$ acquires a large-deviation form \cite{ecru:85,hu:09}. Using the spectral decomposition of the generalized Master operator we can write in the limit of long times $Z=\mathrm{Tr}\, \rho_s(t)\rightarrow e^{t\theta(s)}$ where $\theta(s)$ is the eigenvalue of $\mathcal{W}_s$ with the \emph{largest} real part. The crucial idea behind the thermodynamics of trajectories approach is to interpret $Z$ as a \emph{partition function} and of $\theta(s)$ as a \emph{free-energy}, i.e.\ to ascribe a real physical meaning to these quantities. This allows us to view the ($s$-dependent) mean emission rate $\left<k\right>\!(s)=\left<K\right>\!(s)/t$ of bath quanta as a \emph{dynamical order parameter} which we call the \emph{activity}. The activity can be written as the derivative of the free-energy with respect to the conjugate field $s$: $\left<k\right>\!(s)=-\partial_s \theta(s)$. Likewise, higher moments of the emission statistics are encoded in higher derivatives of $\theta (s)$. Of particular interest here is Mandel's $Q$-parameter, $Q(s) = - \partial^2_s \theta (s) / \partial_s \theta (s) - 1$, which quantifies the deviation of the emission statistics from a Poissonian distribution.

We would like to remark that technically the thermodynamics of trajectories approach bears similarities to that of ``Full Counting Statistics'' (FCS) \cite{lele+:96,lere:04}. However, despite the similarities to FCS we approach the counting problem from a somewhat different angle. We regard the counting field $s$ as the conjugate field to a dynamical order parameter (the event count $K$). This very perspective allows to construct a theoretical framework similar to equilibrium statistical mechanics for the analysis of ensembles of stochastic trajectories. For example, just like in equilibrium statistical mechanics, non-analytical points of the dynamical free-energy $\theta(s)$ will determine transitions between \emph{dynamical} phases \cite{gale:10,gaar+:11,bu:10}. Of particular importance are of course non-analyticities that occur at or very near the physical point $s=0$, as they strongly influence the dynamical behavior of the system.

Before we further illustrate this point by investigating the emission statistics of a particular model system, let us use the approach outlined above to assess possible connections between dynamical and static quantities. In particular, we will show that the activity can directly be related to the expectation value of a static observable. We will also see that, in contrast, the same is in general not true for fluctuations (and higher moments) of the dynamical order parameter.

\subsection{Connection between dynamic and static quantities}\label{sec:connection}
As a consequence of the large deviation principle, the dynamical free-energy  for long times can be written as $\theta(s)=  \partial_t \text{ln}Z = \partial_t \ln \mathrm{Tr}\, \rho_s= (\mathrm{Tr}\, \partial_t\rho_s)/(\mathrm{Tr}\, \rho_s)$. Using the equation of motion for $\rho_s$, we get $\mathrm{Tr}\,\partial_t\rho_s=\mathrm{Tr}\,\mathcal{V}_s(\rho_s)$, where we have exploited that $\mathrm{Tr}\,\mathcal{W}(\rho_s)=0$. Combining these results yields
\begin{equation}
\theta(s) =(e^{-s}-1)\mathrm{Tr}\left(\sum_{\nu} J^\dagger_{\nu} J_{\nu} R(s)\right),
\label{eq:theta}
\end{equation}
where we have introduced the normalized density matrix $R(s) = \rho_s/\text{Tr} \rho_s$, which becomes stationary in the long time limit. Furthermore $R(s)$ coincides with the stationary density matrix of the system at the physical point, $R(s=0) = \rho$. Differentiation of Eq.\ (\ref{eq:theta}) with respect to $s$ reveals the relation between the dynamical order parameters and the expectation value   $\left< \sum_{\nu} J^{\dagger}_{\nu} J_{\nu} \right>$ at $s=0$,
\begin{equation}
 \left<k\right> (0) = -\partial_s\theta(s)|_{s=0}=\mathrm{Tr}\,\sum_{\nu} J^\dagger_{\nu} J_{\nu} \rho. \label{eq:correspondence}
\end{equation}

A further differentiation of Eq.\ (\ref{eq:theta}) shows that this direct connection between dynamical and static quantities does not hold for higher moments. In particular, temporal fluctuation of emission records (quantum jump trajectories) are in general not solely determined by static spatial correlations. Assessing higher moments of the event count distribution, therefore, requires knowledge of the dynamical phases of the system in the vicinity of the physical point, i.e.\ for $s \ne 0$. This can, e.g. been seen when evaluating the $Q$-parameter,
\begin{equation}
Q (0) = -2 \frac{ \mathrm{Tr}\,\sum_{\nu} J^\dagger_{\nu} J_{\nu} \partial_s R(s)|_{s=0}}{\mathrm{Tr}\,\sum_{\nu} J^\dagger_{\nu} J_{\nu} \rho}.
\end{equation}
Calculating $Q$ requires the evaluation of the first derivative of $R(s)$ with respect to the field $s$, i.e. information about $R(s)$ for $s \ne 0$.

This is in fact known from strongly interacting \emph{classical} systems. There, dynamical fluctuations are not necessarily simple manifestations of static fluctuations. A notable example is glasses: there, the thermodynamics can  essentially be trivial, while the corresponding dynamics can be very complex. For example, Refs.\ \cite{gale+:07,heja+:09} show that both idealized lattice models and realistic liquid models, when explored by means of techniques similar to those employed here, display phase transitions in dynamical trajectories which have no static counterpart. To uncover these dynamical transitions it is necessary to go beyond purely static treatments and consider the statistical properties of trajectories.

In the following we will apply the general ideas of the thermodynamics of trajectories approach to study the dynamical behavior of a specific open many-body quantum system.

\section{Dissipative Ising model in a transverse magnetic field}\label{sec:oising}
We consider a system of $N$ spin-$1/2$ particles arranged on the sites of a regular lattice with lattice spacing $a$ and coupled to a thermal bath with zero temperature. The Hamiltonian $H$ is that of a quantum Ising model in a transverse field \cite{sa:08}
\begin{eqnarray}
  H= \Omega \sum_{\nu} S_x^{(\nu)}+ V \sum_{\langle \mu, \nu \rangle} S_z^{(\mu)}S_z^{(\nu)}
    \label{eq:Hamiltonian}
\end{eqnarray}
with $\Omega$ characterizing the transverse field strength and $V$ being the interaction strength between adjacent spins ($\langle\mu,\nu\rangle$ indicates that the sum is only over nearest neighboring pairs).  The Lindblad operators are given by $J_{\nu}=\sqrt{\kappa}S_-^{(\nu)}=\sqrt{\kappa}[S_x^{(\nu)}-iS_y^{(\nu)}]$ with decay rate $\kappa$. This model is sketched in Fig.\ \ref{fig:system}(a).

A natural static order parameter for this system is the magnetization $m=N^{-1}\,\sum_{\nu}\! \langle S_z^{(\nu)} \rangle$, which has also been used to characterize dynamical properties in Refs.\ \cite{leha+:11,leha+:12}. Using Eq.\ (\ref{eq:correspondence}) we immediately see that the static order parameter is indeed proportional to the dynamical order parameter,
\begin{equation}
  \left<k\right> (0) = \kappa N\left[m+\frac{1}{2}\right].
  \label{eq:activity_magnetization}
\end{equation}
The static magnetic susceptibility, however, is not directly related to quadratic fluctuations in the activity, as the former only captures static spatial correlations, while the latter also includes correlations in time.

\subsection{Realization with Rydberg atoms}\label{sec:experiment}
Experimentally the dissipative Ising model can - to a good degree of approximation - be achieved with electronically excited ultracold alkali atoms confined to a lattice. Here, the state $\left|\downarrow\right>$ is identified with the atomic ground state and the state $\left|\uparrow\right>$ is an electronically excited Rydberg nS-state. When two atoms on the $\nu$-th and $\mu$-th lattice site (with position vectors $\mathbf{r}_\nu$ and $\mathbf{r}_\mu$, respectively) are excited simultaneously they interact via a van-der-Waals potential of the form $V_{\nu\mu}=\frac{C_6}{|\mathbf{r}_\nu-\mathbf{r}_\mu|^6} \left|\uparrow\right>_{\nu}\!\left<\uparrow\right|\otimes \left|\uparrow\right>_{\mu}\!\left<\uparrow\right|$. Here $C_6$ is the dispersion coefficient which characterizes the interaction strength. Due to the quick decay of the interaction as a function of the distance one can replace the van-der-Waals potential by a nearest-neighbor interaction \cite{le:11}. When Rydberg states are excited by a laser of Rabi-frequency $\Omega$ and detuning $\Delta$ with respect to the energy difference of the transition $\left|\downarrow\right>\leftrightarrow\left|\uparrow\right>$, the Hamiltonian for the $\nu$-th atom reads $h_{\nu}=\Omega S_x^{(\nu)}+\Delta S_z^{(\nu)}$.  Combining this with the interaction one finds that by choosing $\Delta =-V=-C_6/a^6$ one arrives at Hamiltonian (\ref{eq:Hamiltonian}). The dissipative dynamics is realized naturally as Rydberg states decay radiatively. For $nS$-states of alkali metal atoms this decay takes place predominantly to the lowest $P$-state (on the time-scale of a few microseconds) and subsequently to the ground state. The last decay is very fast (tens of nano-seconds) such that it can be considered instantaneous. Within this approximation we find the dynamics of the system to be governed by the Master operator (\ref{eq:ME}). The activity can then be directly monitored by detecting photons that are emitted during the decay from the lowest $P$-state to the ground state.

\begin{figure}
\centering
\includegraphics[width=\columnwidth]{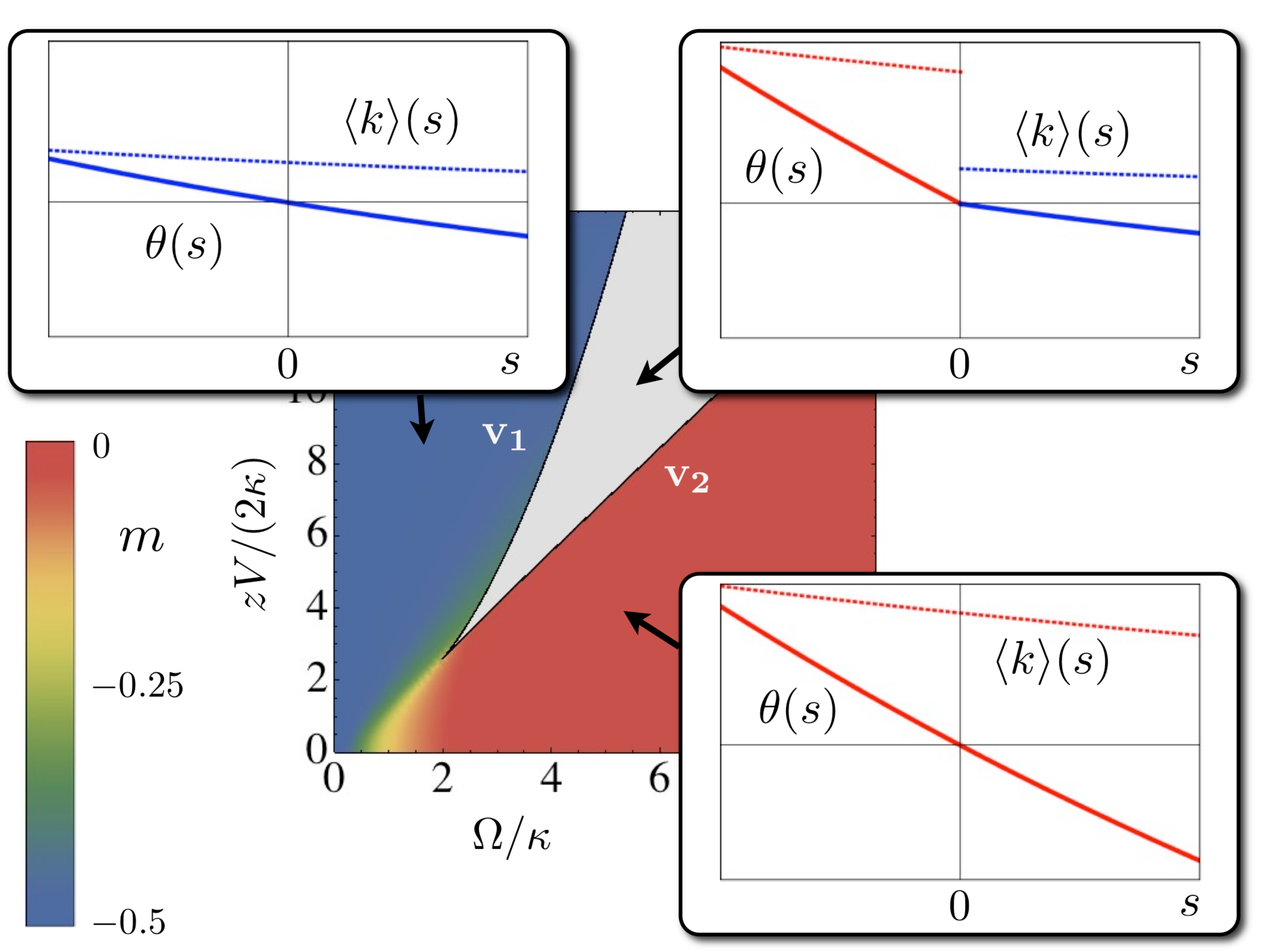}
\caption{(Color online) Mean-field phase diagram. Colored regions have a unique steady state solution. In the grey region, delimited by the two spinodal lines $\mathrm{v}_1$ and $\mathrm{v}_2$ (see text), two steady state solutions with different magnetization exist. These solutions merge at $(\Omega^*; V^*)= (2\kappa; (3\sqrt{3}/z)\kappa)$. The small panels show the behavior of the free-energy $\theta(s)$ (solid) and its first derivative $\left<k\right>\!(s)$ (dotted) in the vicinity of $s=0$. While in the region with unique steady state both functions are smooth, the dynamical order parameter $\left<k\right>\!(s)$ has a jump at $s=0$ for parameter values taken in the grey region. This indicates the existence of a first order dynamical phase transition.}
\label{fig:phases}
\end{figure}

\subsection{Mean-field approximation to statics and dynamics}\label{sec:meanfield}
The static properties of our open Ising model, including the nature of its structural phases and the transitions between them, are given by its stationary density matrix. This is obtained from the eigenstate(s) of the operator $\mathcal{W}$ with zero eigenvalue \cite{brpe:02}. Finding the exact eigenstate(s) is a difficult task, but we can learn much from a mean-field approximation \cite{leha+:11,leha+:12}.

Since the dynamical order parameter is directly proportional to the static one [cf.\ Eq.\ (\ref{eq:activity_magnetization})], determining the activity is equivalent to calculating the magnetization.
In mean-field approximation this can be done in a standard manner by approximating the many-body density matrix $\rho$ as a product of single-particle density matrices, $\rho \approx \bigotimes^N_{\nu=1}\,r^{(\nu)}$ and solving $\mathrm{Tr}_{2...N}\,\mathcal{W}\left(\bigotimes^N_{\nu=1}\,r^{(\nu)}\right)=0$ \cite{bo:98}. The partial traces are taken over the degrees of freedom of all spins but one. Parameterizing the single particle matrices $r^{(\nu)}$ as $r^{(\nu)}=\mathbb{1}/2+\alpha S_x^{(\nu)}+ \beta S_y^{(\nu)}+ m S_z^{(\nu)}$ this procedure leads to the following self-consistency equation for the magnetization $m$:
\begin{eqnarray}
 0 = 8 z^2 V^2 m^3 + 4 z^2 V^2m^2 + 2 (\kappa^2 + 2 \Omega^2) m + \kappa^2,
 \label{eq:selfconsistent}
\end{eqnarray}
where $z$ denotes the coordination number of the lattice. The analysis of this third order polynomial shows that below a critical interaction strength $V^*=(3\sqrt{3}/z)\kappa$ the self-consistency equation (\ref{eq:selfconsistent}) has only one stable real solution. Above $V^*$ one finds, depending on the value of $\Omega$, either one or two stable real solutions. The two spinodal lines \cite{chlu:00} separating the regions of unique and multiple real solutions for $m$ are given by 
\begin{eqnarray*}
\mathrm{v}_1 &\equiv & \frac{zV_1}{2\kappa} \approx \frac{1}{8} \left(\frac{\Omega}{\kappa}+\sqrt{\left(\frac{\Omega}{\kappa}\right)^2 + 2}\right)^2 \\
\mathrm{v}_2 &\equiv & \frac{zV_2}{2\kappa} \approx \frac{1}{\sqrt{2}} \left(\frac{\Omega}{\kappa} +\sqrt{\left( \frac{\Omega}{\kappa} \right)^2 -1}\right), 
\end{eqnarray*}
valid for $V/\kappa,\Omega/\kappa\gg1$.

The corresponding phase diagram is shown in Fig.\ \ref{fig:phases}. In the colored regions there is a unique steady state. In the grey domain two stable solutions exist, one with small magnetization $m_\mathrm{A} \sim 0$ and one with large negative magnetization $m_\mathrm{I}\sim -1/2$. By virtue of Eq.\ (\ref{eq:activity_magnetization}) we can now conclude that the activity $\left<k\right>$ is small (large) in the blue (red) regions. To understand the dynamical behavior of the Ising system in the grey region we have to explore the behavior of $\theta(s)$ in the vicinity of $s=0$.

For this, we expand $\theta(s)$ given in Eq.\ (\ref{eq:theta}) to first order in $s$ and find that $\theta(s) \approx -s \sum_{\nu}\mathrm{Tr} J^{\dagger}_{\nu} J_{\nu} R(0) =-s \langle k \rangle (0) = -s\kappa N \left( m +1/2 \right)$. Taking the magnetization obtained from the mean-field calculation we see that in the colored regions of Fig.\ \ref{fig:phases} we obtain a $\theta(s)$ whose first derivative is smooth (see top left and bottom right panel). In the grey region one has to compare the values of $\theta(s)$ that one obtains for the two solutions $m_\mathrm{A}$ and $m_\mathrm{I}$, choosing the one that maximizes $\theta(s)$, hence
\begin{eqnarray*}
   s>0: &\qquad& \theta(s)=- s \kappa N \left( m_\mathrm{I} +\frac{1}{2} \right) \\
   s<0: &\qquad& \theta(s)=- s \kappa N \left( m_\mathrm{A} +\frac{1}{2} \right) .
\end{eqnarray*}
Since $m_\mathrm{A}\neq m_\mathrm{I}$, the slope of $\theta(s)$ changes at $s=0$, causing a non-analyticity of $\theta(s)$ and a jump of its first derivative, the activity (see top right panel of Fig.\ \ref{fig:phases}).
The grey area in the phase diagram can thus be regarded as a coexistence region of two dynamical phases: an active  and an inactive one. Trajectories in the active phase are dense in quantum jumps and are characterized by a large $\langle k \rangle$; quantum jumps are scarce in trajectories in the inactive phase, which is characterized by a small $\langle k \rangle$. The discontinuity of $\langle k \rangle$ at $s=0$ indicates that the transition between active and inactive phases is of first order.

To illuminate the implications of these results on the emission dynamics of the system let us consider a thermodynamic analogy: a fluid system at the transition point between a high density liquid phase and a low density vapor one. Here, a small change in pressure will either select the liquid or the vapor, which are distinguished by their average specific volumes, confirming that the system is at a point of phase coexistence. Therefore, the ability to vary the pressure yields important information about the nature of the phase transition.  As a further consequence of the first-order coexistence one expects sharp interfaces between the phases. These interfaces are not a property of either phase, but of the fact that there exists a ``surface tension'' between the phases.

In our dynamical case the field $s$ works in exactly the same way as the pressure in our liquid-vapor analogy - it selects (depending on its sign) an active (bright) or inactive (dark)  phase. This allows us to uncover a first order dynamical phase transition and again shows the necessity of studying the system away from $s=0$. Furthermore, we expect to observe sharp interfaces between the dynamical phases. Since quantum jump trajectories of many-body systems live in space and time, the interfaces between distinct dynamical phases can be temporal. The mean-field results, therefore, indicate an intermittent emission pattern, for system parameters chosen from the grey region of the phase diagram shown in Fig.\ \ref{fig:phases}. This region of phase coexistence ends at the point $(\Omega^*; V^*)= (2\kappa; (3\sqrt{3}/z)\kappa)$, reminiscent of the static critical point beyond which liquid-vapor coexistence is possible \cite{chlu:00}.

The mean-field approximation above disregards the finite range of the spin-spin interactions. Such a treatment would be accurate for a fully connected problem (i.e., every spin interacts with all other spins with the same strength), as e.g.\ studied in \cite{leha+:12}. There, one might indeed anticipate strongly collective dynamics. 
However, the spin system studied in this work exhibits \emph{local} interactions, and, therefore, it is not evident \emph{a priori} (particularly in low dimensions) that the collectiveness in photon emission or the intermittence predicted by a mean-field analysis should persist when going beyond this approximation, as we will do in the following.

\begin{figure}
\centering
\includegraphics[width=\columnwidth]{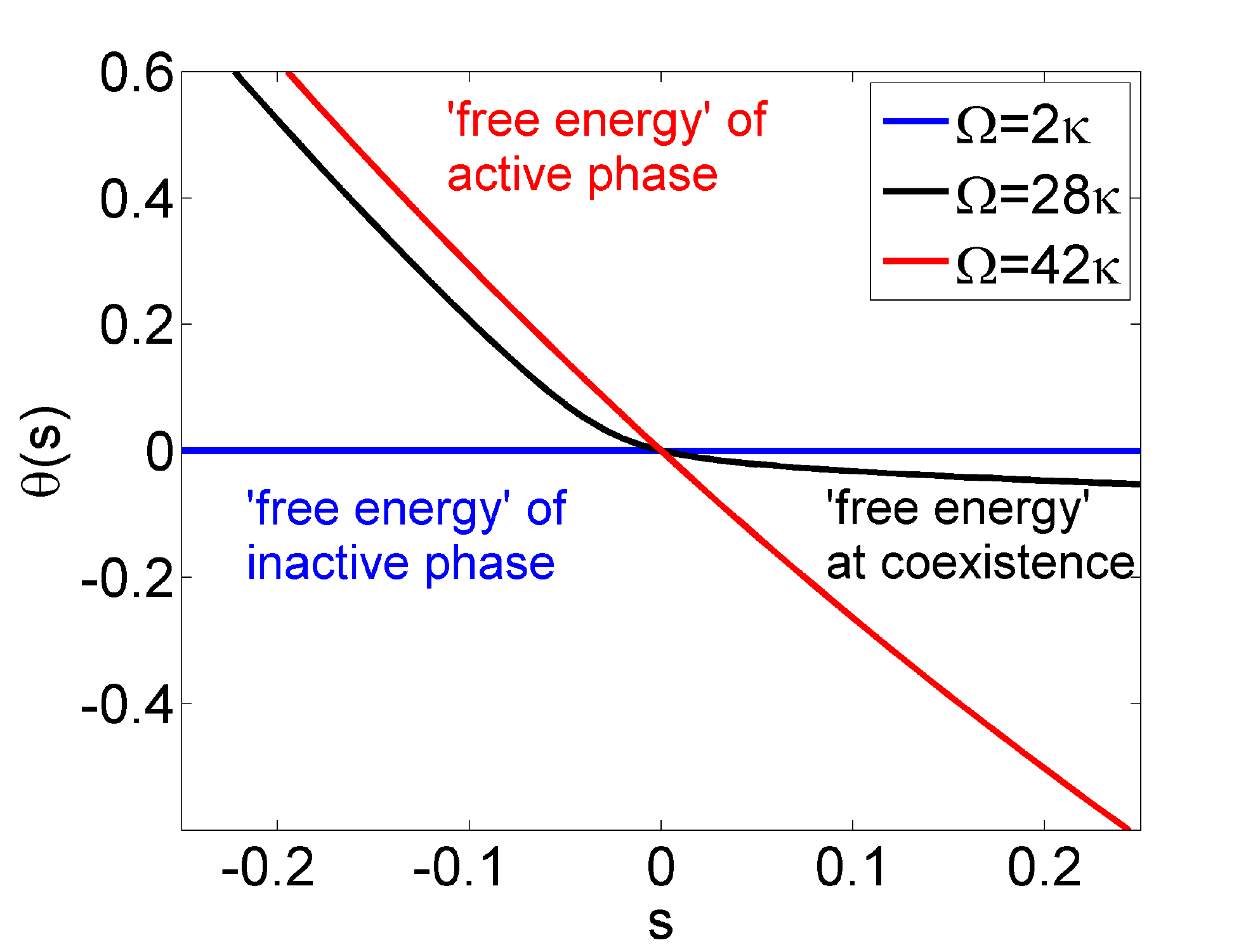}
\caption{
(Color online) Dynamical free-energy $\theta (s)$ determined by exact diagonalization of the generalized Master operator for the one-dimensional, dissipative Ising model in a transverse field for $N=6$ spins with periodic boundary conditions.
The plot shows $\theta (s)$ for $V=100 \kappa$ at three different values of the coupling strength $\Omega$ corresponding to: an active (red) and inactive (blue) dynamical phase and the behavior of the dynamical free-energy under conditions of phase coexistence (black line).
}
\label{fig:theta}
\end{figure}
\begin{figure*}
\centering
\includegraphics[width=0.8\textwidth]{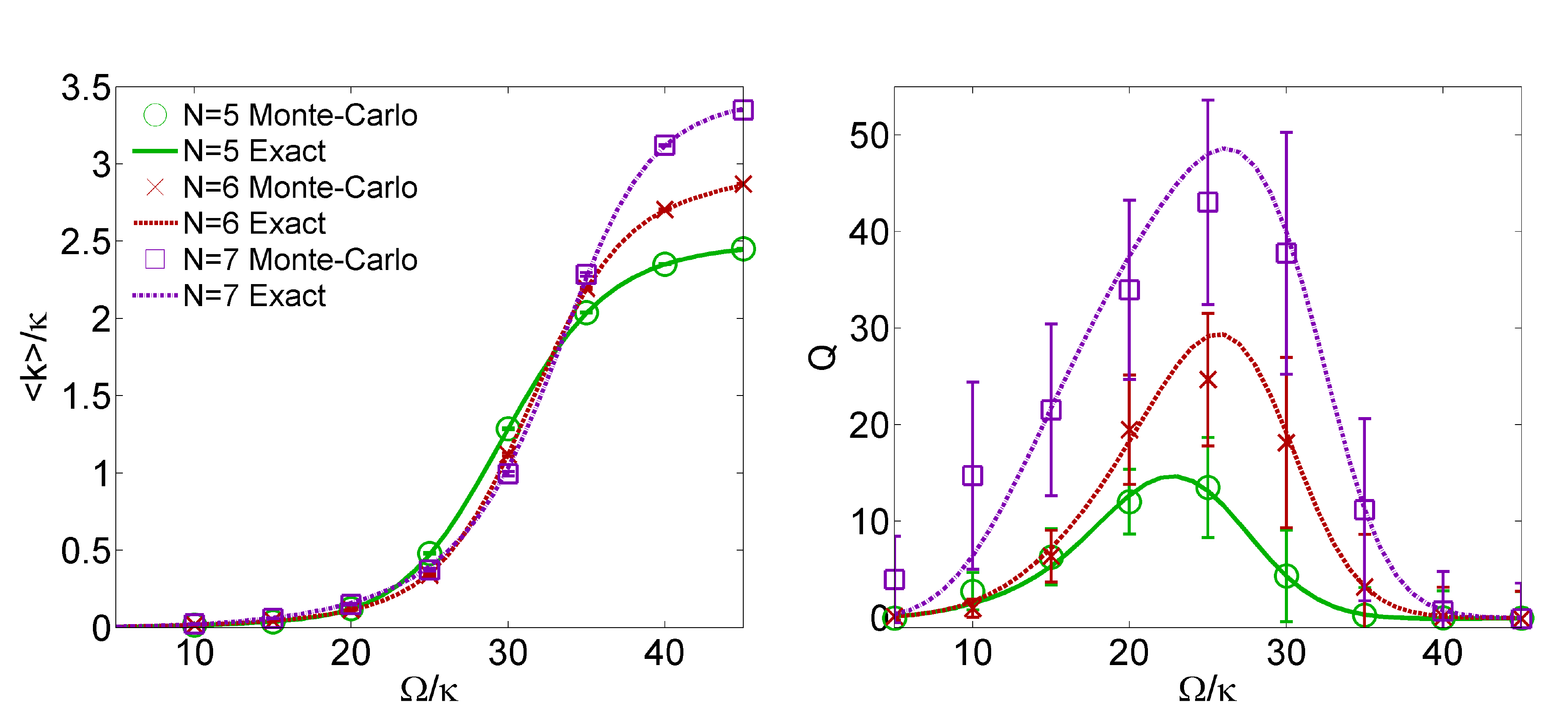}
\caption{
(Color online) Numerically determined activity $\langle k \rangle/\kappa$ (left) and Mandel's $Q$-parameter (right) as a function of $\Omega/\kappa$ for the one-dimensional, dissipative Ising model in a transverse magnetic field for $N=5\dots 7$ sites and periodic boundary conditions. The lines show the results of an exact numerical diagonalization of the generalized Master operator and the symbols the results of Quantum Jump Monte-Carlo simulations (ensembles of 1000 trajectories of length $t=200\kappa^{-1}$ obtained by slicing a much longer one of length $t_\mathrm{max}=2\times10^5\kappa^{-1}$).
}
\label{fig:mandelq}
\end{figure*}

\subsection{Numerical analysis: Exact diagonalization}\label{sec:diagonalization}
To check whether the mean-field treatment yields the qualitatively correct behavior of $\theta(s)$ for our system with short range interactions, we have performed an exact numerical diagonalization of the generalized Master operator for chains up to $N=7$ spins. The largest eigenvalue corresponds to the dynamical free-energy $\theta(s)$ (see e.g. Refs.\ \cite{gale:10,bu:10}).

The results for $\theta(s)$  for $N=6$ spins are shown in Fig.\ \ref{fig:theta} as a function of $s$ for $V=100\kappa$ and different values of $\Omega$. At $\Omega=42\kappa$ (red curve) we find the system in the active phase which is indicated by a large slope at $s=0$. Conversely, at $\Omega=2\kappa$ (blue curve) the activity is small as the slope of $\theta(s)$ at $s=0$ is almost zero. In the region in between [$\Omega=28\kappa$ (black curve)] we find a crossover compatible with a smoothed first order phase transition at $s=0$. This result corroborates the expectations for the large $N$ limit obtained from the mean-field calculation. In fact the curves shown here are a smoothed version of the ones shown in Fig.\ \ref{fig:phases}.

To learn more about the dynamical behavior of the system at the physical point ($s=0$) we have calculated the activity $\langle k \rangle = -\partial_s \theta (s)|_{s=0}$ (lines in left graph of Fig.\ \ref{fig:mandelq}) and Mandel's $Q$-parameter, $Q=-\partial^2_s \theta (s)|_{s=0} /\partial_s \theta (s)|_{s=0} -1$, (lines in right graph of Fig.\ \ref{fig:mandelq}) as a function of the transverse field strength for system sizes of $N=5\dots 7$ spins.  In addition to the results of the exact diagonalization approach, Fig.\ \ref{fig:mandelq} also shows data extracted from Quantum Jump Monte-Carlo simulations (symbols) that will be discussed in the next subsection.

The activity -- being proportional to the magnetization $m$ -- shows the expected behavior as function of the transverse field coupling strength $\Omega$. For small $\Omega$ ($m \approx -1/2$) the system is in an inactive phase with $\langle k \rangle \approx 0$. In contrast, for large $\Omega$ ($m \approx 0$) the system is in an active phase and the activity saturates at $\langle k \rangle/ \kappa = N/2 $ as expected from Eq.\ (\ref{eq:activity_magnetization}). Note that due to the finite size of the system the crossover between these phases is smooth. The $Q$-parameter is close to zero in the inactive (small $\Omega/\kappa$) as well as in active phase (large $\Omega/\kappa$) indicating a Poissonian distribution of the event counts in both dynamical phases. It becomes large and positive in the crossover region, the height of the peak growing with the system size and its position slightly shifting towards larger values of $\Omega/\kappa$. This region is where we anticipate dynamical phase-coexistence in the thermodynamic limit ($t \to \infty$, $\langle K \rangle \to \infty$ while $\langle K \rangle / t = \text{const} < \infty$). This is an indication (but no proof) of the mixing of the two dynamical phases as would be expected in a region of first-order phase coexistence.

To obtain further evidence that the phase mixing scenario does indeed hold, one needs further information about the distribution $p_K$ of the event counts. This can, in principle, be done by determining higher moments of $p_K$ using higher derivatives of $\theta (s)$. This, however, will quickly get intractable in practice, so we will follow an alternative route to show that the picture of the mixing of two dynamical phases does describe the emission dynamics of the system.

\begin{figure*}
\centering
\includegraphics[width=0.75\textwidth]{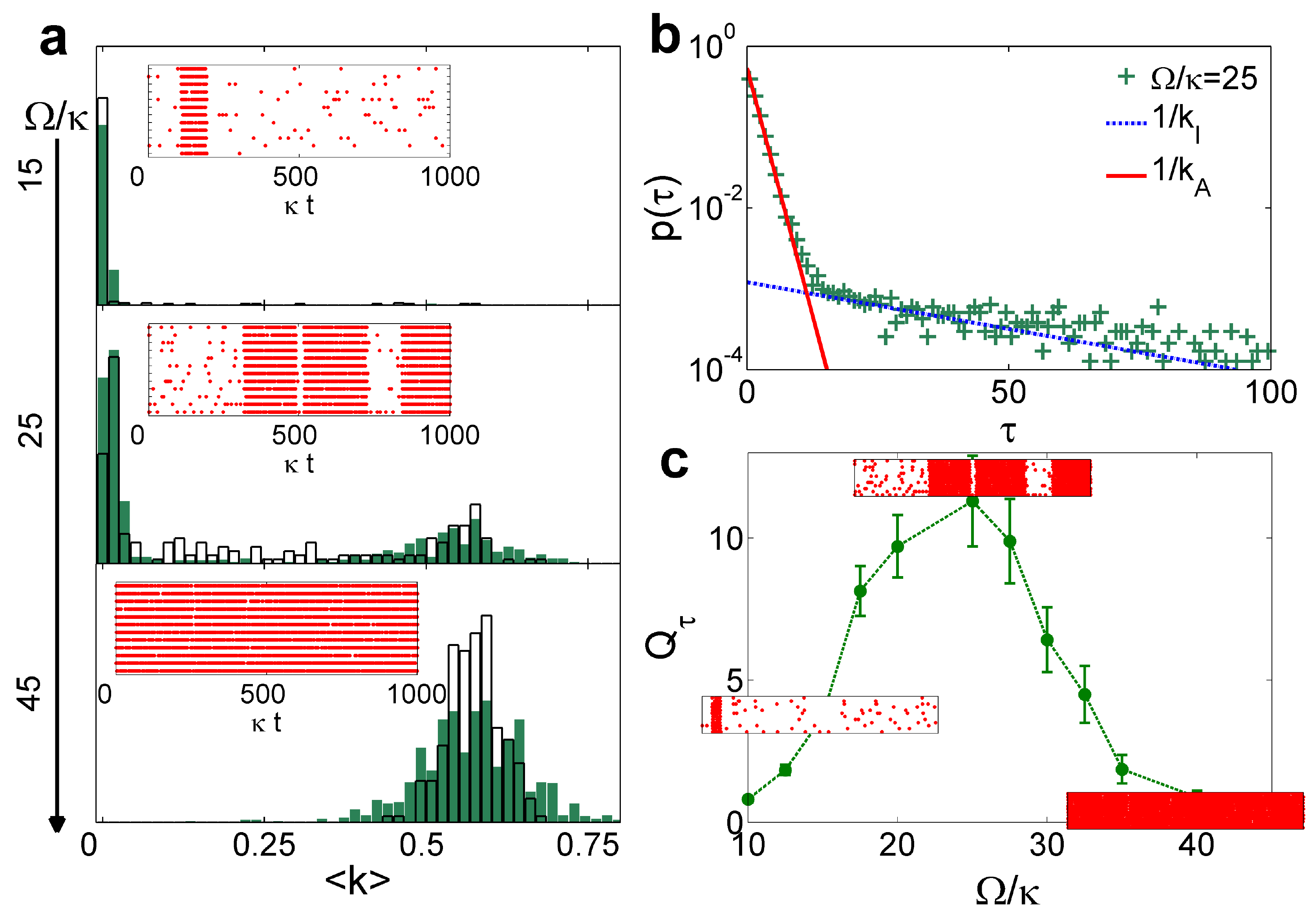}
\caption{(Color online) Analysis of data from Quantum Jump Monte Carlo simulations of a one-dimensional open Ising model in a transverse field with $N=12$ spins ($\kappa=0.1$). (a) Activity distributions of trajectories of length $t=10\kappa^{-1}$ (green) and $50\kappa^{-1}$ (transparent with solid lines), for three different values of $\Omega/\kappa$ for fixed $V=100 \kappa$. These ensembles of trajectories are obtained by slicing longer trajectories of length $t_{\rm max}=10^4 \kappa^{-1}$. We also show sample trajectories at the corresponding conditions, in which the emission events are site resolved. (b) Waiting time distribution $p(\tau)$ for $\Omega=25\kappa$ and $V=100 \kappa$ (crosses in green). The lines represent exponential waiting time distributions with time constants $1/k_\mathrm{A}$ (solid red line) and $1/k_\mathrm{I}$ (dashed blue line). (c) $Q_\tau$ parameter as a function of the transverse magnetic field for $V=100\kappa$ extracted from the waiting time distribution. The insets show samples of typical trajectories in each parameter regime. The maximum of $Q_\tau$ coincides with highly intermittent dynamics.}
\label{fig:analysis}
\end{figure*}

\subsection{Numerical analysis: Quantum Jump Monte-Carlo simulations}\label{sec:montecarlo}
To gain a direct insight into the emission characteristics, we numerically study an unravelling of the full dynamics under the Master operator (\ref{eq:ME}) for spin chains of up to $N=12$ sites using Quantum Jump Monte Carlo (QJMC) simulations \cite{plkn:98}. To connect with potential experiments we can interpret a single quantum jump trajectory of the numerical simulation as a time series of emission events recorded by a detector with temporal and (for illustration purposes here) also spatial resolution, see Fig.\ \ref{fig:system}b. For each set of parameters $(\Omega,V,\kappa)$ this stochastic dynamics generates an ensemble of quantum jump trajectories.

In Fig.\ \ref{fig:mandelq} we compare the mean activity and the $Q$-parameter obtained from QJMC simulations for system sizes of $N = 5\dots 7$ with the results from the exact diagonalization discussed above and find good agreement between them. However, while the activity can be obtained very accurately, the  error bars for $Q$, particularly at intermediate values of $\Omega$, are relatively large. This signifies the strongly fluctuating character of the emission dynamics in the crossover region and shows that in order to capture the dynamical behavior of the system in a possible experiment very long trajectories will be required. Faithfully determining higher moments of the event count distribution will quickly become intractable. However, further information about the counting statistics can still be obtained by resorting to an alternative way to analyze the emission records. To demonstrate this, we have simulated the emission statistics of a chain of $N=12$ spins and analyzed the data by determining the distribution of activities and studying the distribution of the waiting time between quantum jumps (Fig.\ \ref{fig:analysis}).

The mean-field analysis indicates that for certain combinations of $(\Omega,V,\kappa)$ we should encounter a dynamical phase transition of first order. The hallmarks of such a transition would be a bimodal distribution of the activity order parameter and sharp interfaces at coexistence conditions. We generate ensembles of trajectories of length $t$ by slicing much longer trajectories of length $t_{\rm max}\gg t$. The time $t$ can vary, but has to be long enough so that the dynamics within each phase are captured, but not too long compared to the persistence or survival time within each phase which is finite due to the finite size of the system. In Fig.\ \ref{fig:analysis}(a) we show activity histograms at several values of $\Omega/\kappa$ for fixed $V$ and two different values of the time slicing $t=10\kappa^{-1}$ and $50\kappa^{-1}$ (note that the results for both cases are consistent). For small (large) values of $\Omega$ the activity is mostly on the inactive (active) side, as expected from the mean-field analysis. These distributions show bimodality at intermediate values of $\Omega/\kappa$: one peak corresponds to active and the other one to inactive dynamics. When this occurs, the corresponding trajectories are highly intermittent, showing prolonged periods of activity and prolonged periods of inactivity, delimited by sharp temporal interfaces as expected from a first order transition scenario.

Further evidence of this behavior is provided by the waiting time distribution $p(\tau)$ of the emission events [see Fig.\ \ref{fig:analysis}(b)] which shows the existence of two distinct timescales corresponding to the typical waiting times within each phase that govern the dynamics of the system. These two timescales are $1/k_\mathrm{I}$ and $1/k_\mathrm{A}$, which directly relate to the activities of the coexisting inactive and active phases, that are given by $k_\mathrm{I}$ and $k_\mathrm{A}$, respectively.

The results shown in Fig.\ \ref{fig:analysis}(a) and (b) suggest that at the coexistence region the dynamical fluctuations of the open Ising model are dominated by the switching between active and inactive behavior, rather than fluctuations within each of these phases. Disregarding intra-phase fluctuations completely we can approximate the waiting time distribution by the sum of two exponentials, 
\[
p(\tau)=\frac{\alpha k_\mathrm{I}^2 e^{-k_\mathrm{I} \tau}+(1-\alpha) k_\mathrm{A}^2 e^{-k_\mathrm{A} \tau}}{\alpha k_\mathrm{I} + (1-\alpha) k_\mathrm{A}} ,
\]
 i.e. we assume that the active(inactive) dynamics is Poissonian with average number of events per unit time $k_\mathrm{A}$($k_\mathrm{I}$). The lines in Fig.\ \ref{fig:analysis}(b) represent these two contributions, where $k_\mathrm{I}$ and $k_\mathrm{A}$ are extracted from the numerical simulation. The mixing of the two dynamical phases is described by the parameter $\alpha = 0\dots 1$, which determines the point, at which the two lines in Fig.\ \ref{fig:analysis}(b) cross (in the particular case shown $\alpha \approx 0.5$). The dynamical behavior in the intermediate region is strongly mixed and highly non-Poissonian due to the coexistence between two very distinct (close to) Poissonian phases.

Intermittency in the dynamical trajectories corresponds to large fluctuations of the times between quantum jumps. In order to quantify these fluctuations we use the function $Q_\tau = \langle \tau^2 \rangle/\langle \tau \rangle^2-2$ \cite{baju+:04}, where $\langle \tau^{\lambda} \rangle = \int_0^{\infty} \text{d}\tau\; \tau^{\lambda} p(\tau)$ stands for the $\lambda$-th moment of $p(\tau)$. This function reaches its minimum $Q_\tau=-1$ if the waiting time distribution gives rise to a completely regular distribution of the quantum jumps without statistical fluctuations. It assumes the value $Q_\tau=0$ when $p(\tau)$ is an exponential distribution, that is, if the distribution of quantum jumps is Poissonian.  When the fluctuations of times between jumps are large, it assumes a positive value $Q_\tau>0$. The behavior of $Q_\tau$ as a function of $\Omega/\kappa$ is shown in Fig.\ \ref{fig:analysis}(c). It does indeed peak for values of the parameters for which the activity is bimodal and the trajectories intermittent. Far from the coexistence region $Q_\tau$ drops to values close to zero, suggesting that fluctuations within each phase are well approximated by a Poisson process, at least for the long timescales relevant for the discussion here.

\section{Conclusions}\label{sec:conclusions}
We have shown that the concept of a dynamical activity order parameter can be successfully applied to characterize and understand complex dynamical behavior of a many-body quantum system. We have established a general connection between static observables and the dynamical order parameter. For the dissipative Ising model studied here we have identified a first order dynamical phase transition, and showed that coexistence of two dynamical phases gives rise to pronounced intermittency of the bath quanta emission.

\begin{acknowledgements}
This work was funded in part by EPSRC Grant no.  EP/I017828/1 and Leverhulme Trust grant no. F/00114/BG. B.O. also acknowledges funding by Fundaci\'on Ram\'on Areces. C.A. acknowledges support through a Feodor-Lynen Fellowship of the Alexander von Humboldt Foundation.
\end{acknowledgements}


%

\end{document}